# The First Ground Level Enhancement Event of Solar Cycle 24: Direct Observation of Shock Formation and Particle Release Heights

N. Gopalswamy[1], H. Xie[1,2], S. Akiyama[1,2], S. Yashiro[1,2], I. G. Usoskin[3], and J. M. Davila[1]

[1]NASA Goddard Space Flight Center, Greenbelt, Maryland

[2]The Catholic University of America, Washington, DC

[3]Sodankylä Geophysical Observatory (Oulu unit) and Dept. of Physics, University of Oulu, 90014 Oulu, Finland

ABSTRACT

We report on the 2012 May 17 Ground Level Enhancement (GLE) event, which is the first of its kind in Solar Cycle 24. This is the first GLE event to be fully observed close to the surface by the Solar Terrestrial Relations Observatory (STEREO) mission. We determine the coronal mass ejection (CME) height at the start of the associated metric type II radio burst (i.e., shock formation height) as 1.38 Rs (from the Sun center). The CME height at the time of GLE particle release was directly measured from a STEREO image as 2.32 Rs, which agrees well with the estimation from CME kinematics. These heights are consistent with those obtained for cycle-23 GLEs using back-extrapolation. By contrasting the 2012 May 17 GLE with six other non-GLE eruptions from well-connected regions with similar or larger flare size and CME speed, we find that the latitudinal distance from the ecliptic is rather large for the non-GLE events due to a combination of non-radial CME motion and unfavorable solar B0 angle, making the connectivity to Earth poorer. We also find that the coronal environment may play a role in deciding the shock strength.

## 1. Introduction

Ground level enhancement (GLE) in solar energetic particle (SEP) events represents particles accelerated to GeV energies that reach Earth's atmosphere. Typically about a dozen GLE events occur during each solar cycle (Cliver et al. 1982; Cliver 2006; Nitta et al. 2012; Gopalswamy et al. 2012a). GLE events are associated with very fast (~2000 km/s) coronal mass ejections (CMEs) (Gopalswamy et al. 2012a). While there were 16 GLEs during cycle 23, there has been a dearth of GLEs during cycle 24. The Sun is already in its maximum phase (Gopalswamy et al., 2012b), but has produced only one GLE event to date (2012 May 17). There were 2-11 GLEs during the rise phases of cycles 19-23, representing a range from 20% (2 of 10, cycle 19) to 73% (11 of 15, cycle 22) of all GLEs during these cycles.

When and at what height in the corona the GLE particles are released are important questions that have not been fully answered. One of the problems has been the lack of coronagraphic observations close to the Sun. The Large Angle and Spectrometric coronagraph (LASCO) on board the Solar and Heliospheric Observatory (SOHO) had a field of view (FOV) in the range $2.5 - 32$ Rs, except for a couple of events during the years 1997 and 1998. When the GLE-associated CMEs first appeared in the LASCO FOV, the CMEs were already at an average heliocentric distance of ~4.4 Rs. Therefore, one has to back-extrapolate the CME height to the times of shock formation (inferred from associated metric type II radio bursts) and GLE particle release. The 2012 May 17 GLE event is the first one to be fully observed by the Solar Terrestrial Relations Observatory (STEREO) mission with its inner coronagraph (COR1) FOV extending closer to the solar surface (1.4 Rs), so we can determine the CME kinematics and the height of shock formation accurately. STEREO's Extreme Ultra Violet Imager (EUVI) provides additional details of the eruption down to the solar surface. This GLE event also provides the opportunity to verify the CME/shock heights obtained for cycle-23 GLEs using back-extrapolation. Understanding this lone GLE event is important not only for particle acceleration by CME-driven shocks, but also for the weakness of cycle 24.

**2. Observations**

The 2012 May 17 GLE was observed by several polar neutron monitors, Apatity, Oulu and South Pole, with very similar time profiles. Here we use the data from the Oulu Neutron Monitor (ONM), which is part of the World Neutron Monitor Network (Usoskin et al., 2001). ONM detects secondary neutrons produced at the ground level by primary energetic particles with rigidity above ~0.8 GV (the local vertical geomagnetic cutoff rigidity). GLEs are identified as percentage of enhancement of the neutron monitor count above the background. The intensity (corrected for atmospheric pressure) of the 2012 May 17 GLE reached a peak value of 18.6%. Applying various levels of smoothing to the 1-min data, we determined the onset time at Earth to be in the range 01:38 to 01:45 UT with a median value of 01:43 UT, at which time the intensity was at 2% of the peak. Assuming a Parker-spiral length of 1.2 AU, we see that ~1 GeV particles (speed of ~$2.63 \times 10^5$ km/s) take ~11.4 min to reach Earth (assuming scatter-free propagation for the first-arriving particles). Remote-sensing the associated phenomena using coronagraphic, EUV imaging, and radio spectral instruments all involve propagation of electromagnetic signals to Earth (in ~8.3 min), so we normalize the solar particle release (SPR) to the of time these signals (Kahler 1994). Thus, the normalized SPR time (01:40 UT) is 3.1 min earlier than the Earth onset time.

The GLE was associated with an M5.1 flare starting, peaking, and ending at 01:25, 01:47, and 02:14 UT, respectively. A metric type II burst was reported by the Solar Geophysical Data to start at 01:31 UT, which we corrected to 01:32 UT by directly examining the dynamic spectra from Hiraiso, Culgoora, and Learmonth observatories. The STEREO Ahead (STA) and STEREO Behind (STB) spacecraft were leading and trailing Earth at $114^{\circ}.8$ and $117^{\circ}.6$, respectively during the GLE. The GLE source active region (AR 11476) at N11W76 in Earth view corresponds to E39 in STA (disk event) and W194 in STB (backside event) views. STA/COR1 and LASCO/C2 first observed the CME at 01:40 UT and 01:48 UT, respectively.

## 3. CME Kinematics and Height at SPR

We fitted a flux rope to the CME in STEREO and SOHO/LASCO FOV using the Graduated Cylinder Model (Thernisien, 2011), in which the flux rope expands self-similarly with a circular front and conical legs of circular cross section. The flux rope fit gives the CME height-time (h-t) information, width (face-on and edge-on), and the heliographic coordinates of the source region. Figure 1 shows the h-t, speed, and acceleration plots of the 2012 May 17 CME. The CME attained its peak speed (~1997 km/s) at 02:00 UT and acceleration (1.77 km/s$^2$) at 01:36 UT. The acceleration ceased between 01:55 and 02:00 UT, around the time of the soft X-ray peak and peak CME speed. Combining the flare rise time (22 min) and the peak CME speed (1997 km/s), we get the average initial CME acceleration as 1.51 km/s$^2$, consistent with the direct measurements. After attaining the peak speed, the CME slowly decelerated (-6.4 m/s$^2$) due to the drag force.

### 3.1 CME Height at SPR

The STA/COR1 image at 01:40:26 UT (Fig.1b) coincided with the SPR time, so we directly measure the CME height at SPR as 2.32 Rs, consistent with the heights obtained for well-connected events of cycle 23 (Gopalswamy et al. 2012a). The LASCO CME was at 3.57 Rs in the 01:48 UT image, which can be back-extrapolated to the SPR time as 2.2 Rs in agreement with the COR1 measurement. This was possible because the CME reached a quasi-constant speed of 1997 km/s around this time.

**3.2 Height of Shock Formation**

The CME height at the metric type II burst onset is taken as the height of shock formation (Gopalswamy et al. 2012a,b). In fast events like this, the separation between the flux rope and the shock is generally small, so we take the CME height to be the shock height. The STA/EUVI image at 01:30:56 UT (near the type II burst onset) shows the EUV disturbance associated with the CME. A circle fit to the outermost parts of this disturbance has a radius of 0.31 Rs (Fig. 2a), which represents the CME height above the solar surface for an initial spherical expansion (Gopalswamy et al. 2013). With a slight extrapolation using the local speed (698 km/s) and acceleration (1.77 km/s$^2$), we get the CME height at type II onset (01:32 UT) as 1.38 Rs (from the Sun center). For cycle-23 GLEs, the CME height at SPR ($h_s$) showed a parabolic dependence on the source longitude ($\lambda$): $h_s = 2.55 + [(\lambda-51)/35]^2$ (see Fig.2b and Reames, 2009). The cycle-24 GLE agrees with this: for $\lambda = 76°$, the parabola gives $h_s$=3.06 Rs, similar to the measured value. The CME height at type II onset ($h_{II}$=1.38 Rs) is also similar to the average $h_{II}$ (1.53 Rs) for cycle-23 GLEs (the deviation is only 9.8%).

**4. What is Special about the 2012 May 17 GLE?**

The flare size (M5.1) of the 2012 May 17 GLE is rather small for a GLE event. There were 41 ≥M5 flares during cycle 24 (until the end of 2012), eight of which were well connected (longitude range: W55-W88 similar to the GLE event) (see Table 1). All flares had CME association, except for the confined flare on 2011 February 18. Flux-rope (FR) fit yielded CME speeds (1997 to 2905 km/s) and initial accelerations ($a_i$: 0.95 to 2.31 km/s$^2$) that are typical of GLE events (Gopalswamy et al. 2012a). These speeds are remarkably high for non-GLE events. The edge-on and face-on half widths (w) of FRs were similar except for the relatively narrow CME on 2012 July 8. All the CMEs were associated with large SEP events (proton intensity $I_p \geq$ 10 pfu in the >10 MeV GOES channel with 1 pfu (particle flux unit) = 1 particle.cm$^{-2}$.s$^{-1}$.sr$^{-1}$) in the range 26 to 800 pfu). All the CMEs had type II radio bursts extending from metric to kilometric domains suggesting that the CMEs were driving shocks far into the interplanetary medium.

Table 1. Well-connected major flares ($\geq$ M5.0) and CMEs in cycle 24.

| Flare date & Time | AR # | Flare Imp. | Flare Loc. | FR Loc. | B0 (Deg) | Final Lat. | CME Time | $V_{max}$ (km/s) | $a_i$ (km/s$^2$) | w (deg) | $I_p$ (pfu) | $E_{max}$ (MeV) |
|---|---|---|---|---|---|---|---|---|---|---|---|---|
| 2011/02/18 09:55 | 11158 | M6.6 | S21W55 | ---- | -6.92 | ---- | ---- | ---- | ---- | ---- | ---- | ---- |
| 2011/08/09 07:48 | 11263 | X6.9 | N17W69 | N08W68 | +6.33 | N01.7 | 08:12 | 2496 | 1.56 | 38/47 | 26 | 465 |
| 2012/01/27 17:37 | 11402 | X1.7 | N27W71 | N35W78 | -5.59 | N40.6 | 18:27 | 2755 | 1.48 | 41/48 | 800 | 605 |
| 2012/03/13 17:12 | 11429 | M7.9 | N17W66 | N21W52 | -7.20 | N28.2 | 17:35 | 2333 | 1.13 | 33/49 | 500 | 375 |
| 2012/05/17 01:25 | 11476 | M5.1 | N11W76 | S07W76 | -2.41 | S04.6 | 01:48 | 1997 | 1.77 | 33/46 | 255 | >700 |
| 2012/07/06 23:01 | 11515 | X1.1 | S13W59 | S29W62 | +3.44 | S32.4 | 23:12 | 2464 | 1.21 | 29/48 | 25 | 375 |
| 2012/07/08 16:23 | 11515 | M6.9 | S17W74 | S34W88 | +3.65 | S37.7 | 16:48 | 2905 | 1.31 | 16/28 | 18 | 375 |
| 2012/07/19 04:17 | 11520 | M7.7 | S13W88 | S15W88 | +4.73 | S19.7 | 05:36 | 2048 | 0.95 | 31/49 | 70 | 375 |

The source locations from the fit (FR Loc.) of non-GLE CMEs are significantly different from the flare locations (Flare Loc.) implying non-radial CME motion either inherent to the active region or due to deflection by coronal holes (Gopalswamy et al., 2009; Xie, Gopalswamy and St.Cyr, 2012; Mäkelä et al., 2012). The GLE CME also moved non-radially, but towards the ecliptic (FR: S07W76, flare: N11W76 - see Fig. 2b). With one exception (2011 August 09), the latitudes of non-GLE CMEs increased by 2° to 18°. Correcting for the B0 angle further increased the latitudinal distances from the ecliptic ($\theta_e$): S19.7 to S37.7 and N28.2 to N40.6. On the other hand, $\theta_e$ of the GLE and 2011 August 09 events decreased to S4.6 and N1.7, respectively. Figure 3a shows that $\theta_e$ is higher for the non-GLE events in Table 1 compared to that of cycle-23 GLEs (FR locations corrected for B0 angle). For cycle-23 GLEs $<\theta_e> = 14°.1$ (13°.1 if the 1998 August 24 GLE is excluded because it had no CME data to fit a flux rope). For the five non-GLEs in Table 1 $<\theta_e> = 26°.7$ (31°.7 when the 2011 August 9 event is excluded), which is two times higher than $<\theta_e>$ of cycle-23 GLEs. For cycle-23 limb GLEs $\theta_e$ ranged from 2°.1 to 19°.3 (average 9°.3) compared to 19°.7 to 40°.6 (average 31°.7) for the non-GLEs in Table 1. The two sets have no overlap, suggesting poor magnetic connectivity to Earth for the non-GLE events. Therefore, we cannot rule out the possibility of high energy particle production in some of the CMEs in Table 1, but the particles might not have reached Earth due to poor connectivity.

The well-connected CME on 2012 May 17 CME resulted in a GLE, but not the 2011 August 9 CME: even though the latter had a bigger flare (X6.9), faster CME (2496 km/s), and better connectivity (N1.7), the mean energy ($E_{max}$) of the highest energy channel with a significant SEP intensity was 465 MeV (last column of Table 1). Figures 3b,c,d show the highest energy channel with significant SEP intensity: 350-420 MeV (2011 August 9), 510-700 MeV (2012 January 27), and >700 MeV (2012 May 17). The gradual rise of the SEP intensity in the 2012 January 27 event does indicate poor connectivity, consistent with high $\theta_e$ (40°.6). We speculate that this would have been a GLE if the source were closer to the ecliptic. As for the 2011 August 11

CME, the corona ahead of it appears dimmer than that ahead of the GLE CME (see Fig. 4). Dimmer corona means lower density and higher Alfven speed, and hence weaker shock (Gopalswamy et al., 2003). Previous studies based on the ability of CMEs to produce type II radio bursts have shown that the coronal Alfven speed can be as high as 1500 km/s and can vary by a factor of 4 (Gopalswamy et al., 2008a,b). The flux rope part of the 2011 August 9 event is rather small and the shock standoff distance is relatively large, which is another signature of a weak shock.

All CMEs in Table 1 were associated with large SEP events, so the lower-energy particles did have access to field lines connecting to Earth irrespective of the large ecliptic distance of the sources (see also Dalla and Agueda, 2010). Shock flanks reaching the ecliptic must have produced these lower-energy particles. A further inference is that GLEs are produced around the shock nose, where the shock is the strongest. Thus, the CME propagation direction and the ecliptic distance of the source region are additional parameters that introduce variability in the GLE and SEP events in addition to the coronal environment (preceding CMEs, seed particles, and ambient Alfven speed). It must be noted that the 2012 May 17 CME was preceded by a hot ejecta (>6 MK) about 40 minutes before from the same active region with a speed of 70 km/s revealed by SDO images. However, it was overtaken by the main CME before the shock formation, so it is not clear if it influenced the production of GLE particles.

## 5. Summary and Conclusions

We analyzed the first and only GLE event during solar cycle 24 (as of this writing) that was fully observed by the STEREO mission, providing accurate CME kinematics. One of the striking aspects of this GLE is its association with a moderate flare (M5.1), but the CME itself was very fast (typical of GLE associated CMEs). The flare size is smaller than that in all cycle-23 GLEs and larger than that in only two pre-SOHO GLEs, which occurred when the pre-event SEP backgrounds were elevated due to preceding activity (Cliver, 2006). The 2012 May 17 event further suggests that the flare size is not a good indicator of GLE production because there were half a dozen well-connected eruptions with larger flare size and faster CMEs during cycle 24 that lacked GLEs. We attribute the lack of GLEs to the fact that the CME noses were too far from the ecliptic indicting poor magnetic connectivity to Earth. Normally, source longitudes are considered important for magnetic connectivity to Earth; we find that the source latitudes are also important.

The main conclusions of this study are:

(i) The CME heights at shock formation and SPR are 1.38 and 2.32 Rs, respectively. The time available for the production of GeV particles is ~8 min, as in cycle-23 GLEs.

(ii) The CME height at SPR ($h_s$) agrees with the parabolic dependence on the source longitude ($\lambda$): $h_s = 2.55 + [(\lambda-51)/35]^2$ derived for cycle-23 GLEs. The direct measurement of $h_s$ also validates the indirect methods employed for the cycle-23 GLEs.

(iii) With one exception, the cycle-24 non-GLE CMEs showed deflection away from the ecliptic rendering them poorly connected to Earth. This result suggests that the ecliptic distance to the shock nose is also an important parameter that decides whether an eruption results in GLE. In the case of the lone exception, the eruption was very well connected, but the ambient corona appears tenuous indicating a weaker shock, consistent with the larger standoff distance observed in this event.

(iv) The B0 angle can also worsen or improve the magnetic connectivity to Earth depending on the time of the year.

STEREO is a mission in NASA's Solar Terrestrial Probes program. SOHO is a project of international collaboration between ESA and NASA. Work supported by NASA/LWS program. Oulu NM data are available at http://cosmicrays.oulu.fi.

**References**

Cliver, E. W. 2006, ApJ, 639, 1206

Cliver, E. W., Kahler, S. W., Shea, M. A., Smart, D. F. 1982, ApJ, 260, 362

Dalla, S., Agueda, N. 2010, Role of latitude of source region in Solar Energetic Particle events, in SOLAR WIND TEN. AIP Conference Proceedings, Volume 679, pp. 613-616

Gopalswamy, N. 2012, Energetic particle and other space weather events of solar cycle 24, in SPACE WEATHER: THE SPACE RADIATION ENVIRONMENT. AIP Conference Proceedings, Volume 1500, pp. 14-19

Gopalswamy, N., Yashiro, S., Michalek, G. et al. 2003, Effect of CME Interactions on the Production of Solar Energetic Particles, in SOLAR WIND TEN. AIP Conference Proceedings, Volume 679, pp. 608-611

Gopalswamy, N., Yashiro, S., Xie, H. et al. 2008a, ApJ, 674, 560

Gopalswamy, N., Yashiro, S., Akiyama, S. et al. 2008b, Ann. Geophys., 26, 1

Gopalswamy, N., Mäkelä, P., Xie, H., Akiyama, S., Yashiro, S. 2009, JGR, 114, A00A22

Gopalswamy, N., Xie, H., Yashiro, S., Akiyama, S., Mäkelä, P., Usoskin, I. G. 2012a, SSRv, 171, 23

Gopalswamy, N., Yashiro, S., Mäkelä, P., Michalek, G., Shibasaki, K., & Hathaway, D. H. 2012b, ApJ, 750, L42

Gopalswamy, N., Xie, H., Mäkelä, P., et al. 2013, J. Adv. Space Res., DOI: 10.1016/j.asr.2013.01.006

Kahler, S. W. 1994, ApJ, 428, 837

Mäkelä, P., Gopalswamy, N., Xie, H., Mohamed, A. A., Akiyama, S., Yashiro, S. 2012, Solar Phys., online first, DOI: 10.1007/s11207-012-0211-6

Nitta, N. V., Liu, Y., DeRosa, M. L., Nightingale, R.W. 2012, SSRv, 171, 61

Reames, D. V. 2009, ApJ, 706, 844

Thernisien, A. 2011, ApJS, 194, 33

Usoskin I. G. Mursula K., Kangas, J. & Gvozdevsky G. 2001, On-Line Database of Cosmic Ray Intensities, Proceedings of ICRC, Hamburg, 2001, p. 3842

Xie, H., Gopalswamy, N. St. Cyr, O. C. 2012, Solar Phys., online first, DOI: 10.1007/s11207-012-0209-0

**Figure Captions**

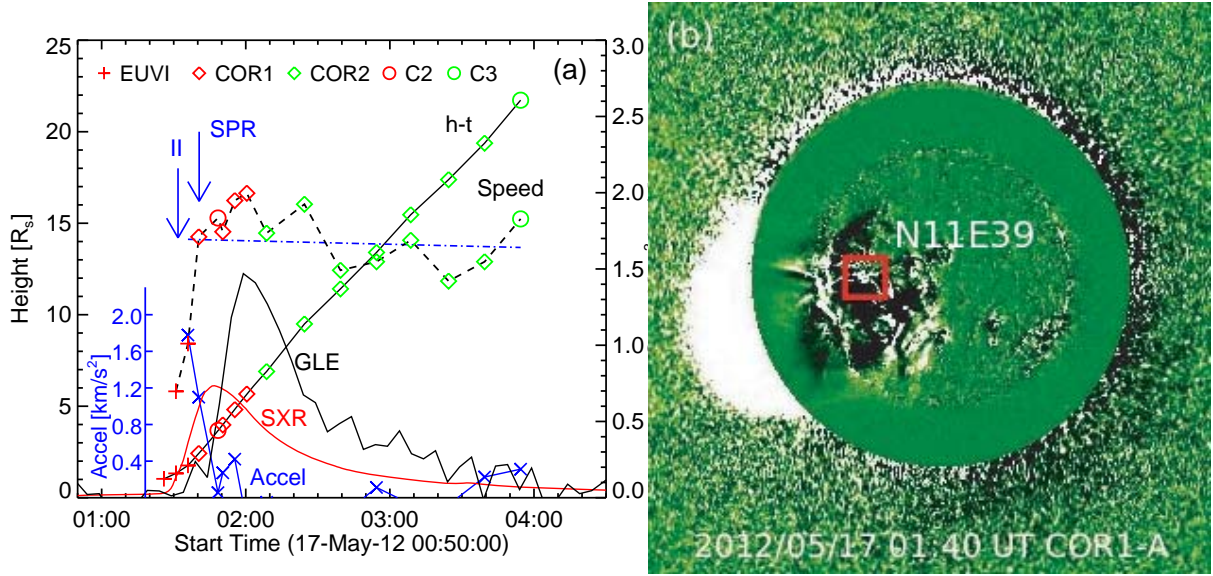

Figure 1. (a) Height-time (h-t), speed, and acceleration plots of the 2012 May 17 CME obtained from STEREO COR1, COR2, EUVI and LASCO C2, C3 data. The line through the speed points indicates a slow deceleration. The GOES soft X-ray flux and GLE intensity (5-min resolution) are in arbitrary units. SPR and metric type II burst onsets are marked by vertical arrows. GLE data points are shifted to the right by 3 minutes assuming propagation along a 1.2-AU Parker spiral. (b) STA/COR1 image at 01:40 UT showing the CME at SPR. A simultaneous STA/EUVI difference image is superposed with the flare location is indicated.

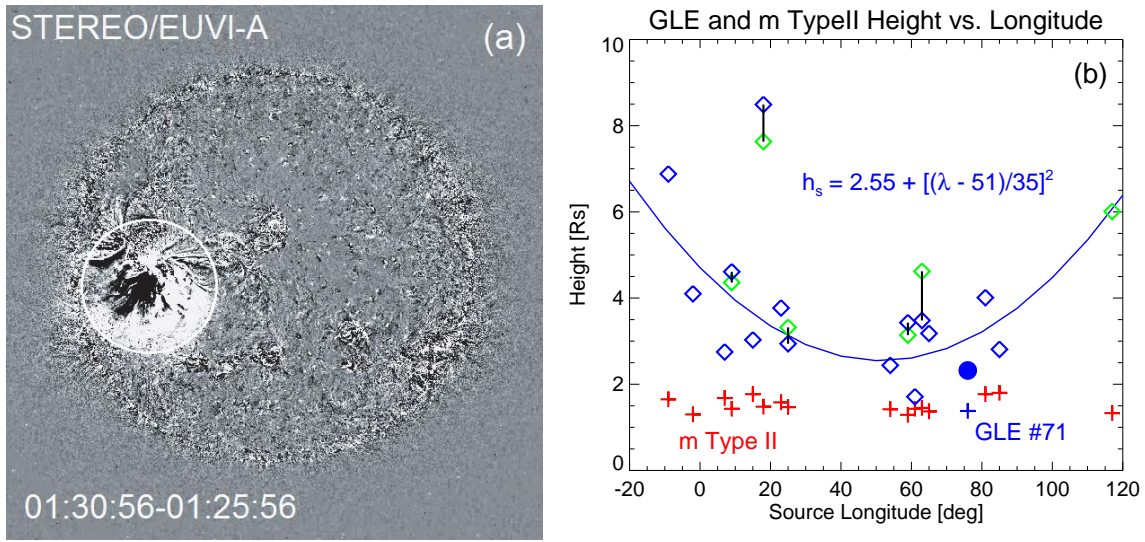

Figure 2. (a) A circle fit to the EUV disturbance in the STA/EUVI image at 01:30:56 UT. The times of STEREO images are corrected to account for the light travel time from the Sun to Earth and STA. (b) The CME height at SPR and type II onsets (filled circle and blue 'plus', respectively) of the 2012 May 17 GLE compared with the corresponding values for the cycle-23 GLEs. The parabola is a fit to the cycle-23 CME heights at SPR (diamonds) as a function of the source longitudes ($\lambda$): $h_s = 2.55 + [(\lambda-51)/35]^2$. The green and blue diamonds indicate heights determined from linear and quadratic extrapolations.

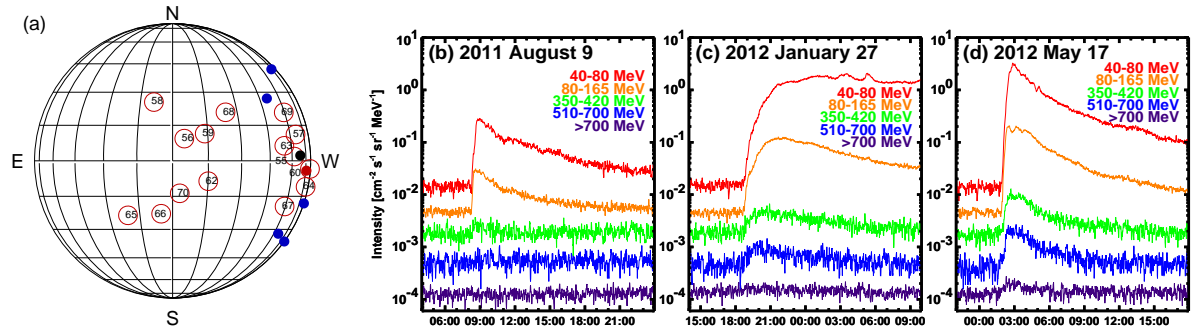

Figure 3. (a) Source locations of CMEs in Table 1 (filled circles) compared with those of cycle-23 GLE CMEs (open circles with GLE numbers). The CMEs on 2012 May 17 (GLE) and 2011 August 09 are distinguished by red and dark colors, respectively. The sources of cycle-23 GLEs were obtained from flux-rope fitting and correcting for solar B0 angle. (b-d) Proton fluxes for three events in Table 1 with SEPs observed in three highest-energy GOES13 channels: 2011 August 9 (b), 2012 January 27 (c), and 2012 May 17 GLE (d) events. (SEP data from GOES 13: http://satdat.ngdc.noaa.gov/sem/goes/data/new_avg/).

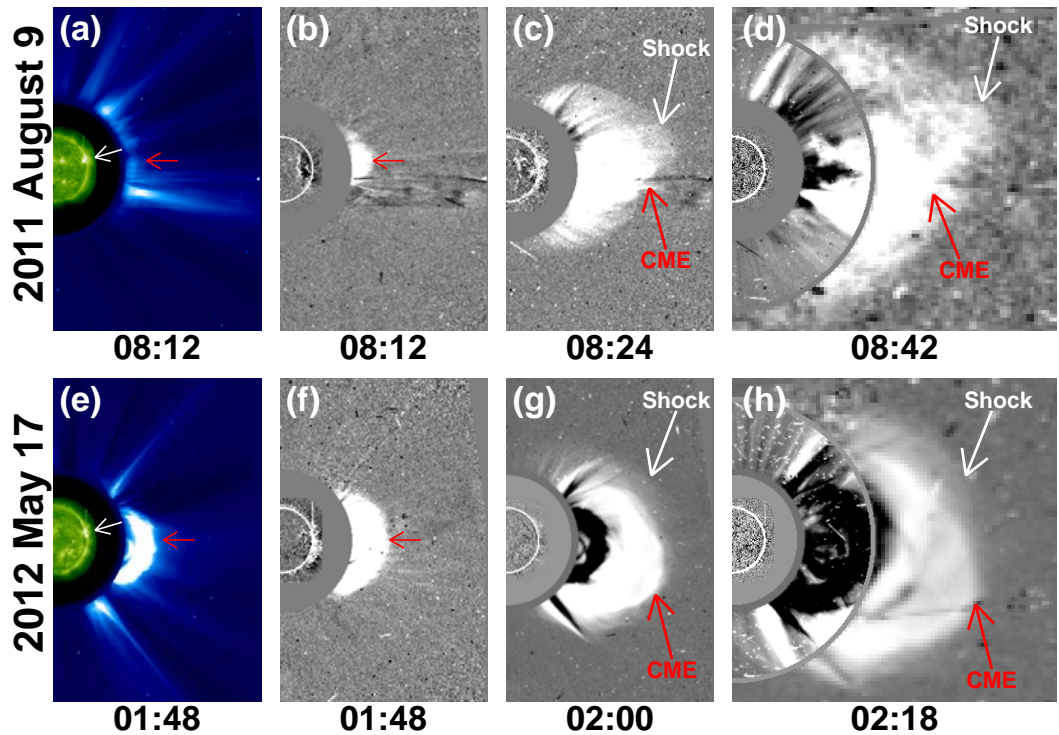

Figure 4. Evolution of the 2011 August 9 (top) and 2012 May 17 (bottom) CMEs. The first column (a,e) gives the direct images from SOHO/LASCO showing the two CMEs at their first appearance. The SDO/AIA 193 Å images show the solar sources. The LASCO images show that the ambient corona ahead of the 2011 August 9 CME is relatively dim and that the CME is also faint. The second column (b,f) shows the difference images of the CMEs at their first appearance. Arrows point to the flare site (white) and CME nose (red). The slightly larger size of the CME in the difference image is due to the fact that the leading edge includes the shock structure. In (c,g) and (d,h) the main body (CME) and the shock structure are indicated by arrows. In the last column (d,h), C2 images are superposed inside the C3 occulting disk. The large gray circle is the outer edge of the C3 occulting disk. The 2012 May 17 CME remains bright with its shock structure close to the CME (g,h). In the 2011 August 9 CME, the main body is somewhat smaller and the shock structure is more extended (c,d), which is a sign of weak shock.